\documentclass[fleqn,usenatbib]{mnras}

\usepackage{newtxtext,newtxmath}

\usepackage[T1]{fontenc}
\usepackage{ae,aecompl}

\usepackage{graphicx}	
\usepackage{amsmath}	
\usepackage{amssymb}	
\usepackage{epsfig}     

\title[Gravitational Waves From Ultra Short Period Exoplanets]{Gravitational Waves  From Ultra Short Period Exoplanets}

\author[Cunha, Silva \& Lima]{
J. V. Cunha,$^{1}$\thanks{E-mail: jvital.cunha@ect.ufrn.br}
F. E. Silva,$^{1}$\thanks{E-mail: edsonsilva@ect.ufrn.br}
J. A. S. Lima$^{2}$\thanks{E-mail: jas.lima@iag.usp.br}
\\
$^{1}$Escola de Ci\^encias e Tecnologia, UFRN, 59078-970, RN, Brazil\\
$^{2}$Departamento de Astronomia, Universidade de S\~ao Paulo, 05508-900 SP, Brazil}

\date{Accepted XXX. Received YYY; in original form ZZZ}

\pubyear{2017}

\begin{document}
\label{firstpage}
\pagerange{\pageref{firstpage}--\pageref{lastpage}}
\maketitle
\begin{abstract}
In the last two decades, thousands of extrasolar planets were discovered  based on different  observational techniques, and their number must increase substantially in virtue of the ongoing and  near-future approved missions and facilities. It is shown that interesting signatures of binary systems from nearby exoplanets and their parent stars can also be obtained measuring the pattern of gravitational waves that will be made available by the new generation of detectors including the space-based LISA (Laser Interferometer Space Antenna) observatory.  As an example, a subset of exoplanets with extremely short periods (less than 80 min) is discussed.  All of them have gravitational luminosity,  $L_{GW} \sim 10^{30}$ $erg/s$, strain $h \sim 10^{-22}$, frequencies $f_{gw} > 10^{-4}$Hz, and, as such, are  within the standard sensitivity curve of LISA. Our analysis suggests that the emitted  gravitational wave pattern may also provide an efficient tool to discover ultra short period exoplanets. 
\end{abstract}

\section{Introduction}
Recently, the LIGO and Virgo collaborations have directly detected gravitational waves ({GWs}) from merging pairs of black holes and neutron stars thereby opening a new observational  window to astrophysics and cosmology (Abott {\it et al.} 2016a, 2016b, 2017). This successful centenary prediction  of general relativity is now pressing for world-wide observational efforts involving the next generation of space-based and ground-based detectors of {GWs}, planned to ``listen" unprobed frequency bands from a large variety of sources.  

The recent detection of {GWs} is the result of a long process. Implicitly, it also suggests that  the existence of  numerous sources with different characteristic frequencies is quite probable. Currently, we are just starting the exciting work of studying and characterizing the basic properties of GWs and their potential sources regardless of its cosmic abundance or even whether they are nearby or very far from the solar system. 

It is also usually believed that the main observable candidates  to GW observations are ultracompact binary black hole and neutron star systems, rotating black holes and supernovas, that is, very massive and extremely compact systems because these violent systems or events can generate gravitational waves intense enough to be detected even at very great distances (Maggiore 2008). However, since these sources are rather exotic it seems interesting to investigate the potentialities of nearby and less extreme sources. In principle, the small distance from nearest sources may compensate  their relatively low intensity thereby providing important targets for a continued observation.

In this concern, an increasing number of exoplanetary systems  were observed in the last few years, some of them composed by super Jupiters with orbits very near to their parent stars. Such galactic system have recurrently been reported and are much closer to the solar system than known pulsars and  binaries of black holes or neutron stars (the only identified sources up to now).  

Beyond the expected emission from the orbital motion, usually treated in the quadrupole approximation,  the gravitational wave intensity from exoplanets can also be amplified from gravitational interaction, as for instance, through the resonant excitation of oscillating modes from  central stars (Kajima 1987; Berti \& Ferrari 2001).  

On the other hand, the already observed population of exoplanets can be extrapolated to obtain the abundance of such objects in  the Galaxy  and even in all the Universe. Recent estimates are  claiming  the existence  of at least one planet per star thereby leading to nearly two hundred billions of exoplanets in the Milky Way (Cassan et al. 2012).  Some authors  have also predicted the collective contribution of exoplanets appearing as a stochastic GW background (SGWB) from the Milky Way and also from the whole Universe, a result relevant to space-based instruments like LISA. In principle, such an approach may provide near future  an interesting window to probe the cosmic abundance of such objects (Ain, Kastha \& Mitra 2015). Here, instead to analyze the collective emission forming the SGWB, a different strategy it will be adopted. 

Basically, we are interested on a special class of exoplanetary systems already identified in the very recent literature. This happens because these Galactic GW sources are very near to us and orbiting  their parent star with ultra short periods, say, of the order of one hour or less. Further, there is  also a growing interest of the community with several large international consortia pointing to an increasing rate of known systems. Thus, it is reasonable to expect that a part of them must have ultra  short periods. 

The gravitational luminosity of such systems can be even greater than their electromagnetic counterpart, an aspect that may have a strong influence on the so-called habitable planetary zone. Thus, since they are not far away, we believe that such systems are excellent candidates for a new class of targets whose pattern of GWs  are not only different from the ones presented by extreme events (like the ringdown of binary black holes) but  also have  the advantage that the emitted GWs can be continuously detected.

In this Letter we show that exoplanetary systems are also interesting sources of gravitational waves for the next generation of detectors. For a subset of extra solar planets orbiting their parent stars with extremely short periods, we calculate the local GW luminosity,  the strain  and the conditions for detecting the emitted pattern of gravitational waves. As we shall see, they are  within the standard sensitivity curve of LISA and other planned instruments. Reciprocally, we also argue that the GW astronomy may provide a complementary window  to identify extrasolar planets, a new procedure that  may be  even  more efficient than the available  methods (like the transit time) based on optical instruments and techniques. 

\section{Basic Equations}\label{sec:Basic Equations}

The gravitational quadrupole emission from two orbiting celestial massive bodies like an extrasolar system formed by a star and  a given planet can be precisely calculated (Peters \& Mathews 1963; Maggiore 2008). If the system have masses $m_1$ and $m_2$  moving in Keplerian orbits around the center of mass with an effective semi-major axis $a$ and eccentricity $e$, the total average gravitational emission over one period of the elliptical motion reads: 
\begin{equation}\label{LGW}
L_{GW}=\frac{32}{5}\frac{G^4}{c^5}\frac{m_1^2m_2^2(m_1+m_2)}{a^5{(1-e^2)^{7/2}}}\left(1+\frac{73}{24}e^2+\frac{37}{96}e^4\right) ,
\end{equation}
which is the same equation used to calculate the gravitational luminosity emitted from ultracompact binary systems (Maggiore 2008; Postnov and Yungelson 2014). For a source of absolute luminosity $L_{GW}$, the flux or apparent luminosity ($\ell$) received on Earth is $\ell = L_{GW}/4 \pi d^2$.

It is also widely known that the emitted radiation in the eccentric case is no longer monochromatic. In general, the absolute  GW luminosity in the $n^\mathrm{th}$ harmonic, at a frequency $f=n \omega / \pi$, is given by
\begin{equation}
L^{(n)}_{GW} \ = \ \frac{32}{5}\frac{G^4}{c^5}\frac{m_{1}^2 m_2^2(m_1+m_2)}{a^5}B(n,e) \, ,
\label{eq:Power}
\end{equation}
where,
\begin{equation}
\begin{split}
B(n,e) & = \frac{n^4}{32}\bigg(\big[ J_{n-2}(ne)-2eJ_{n-1}(ne) \ + \\
\frac{2}{n} J_n(ne) & + 2eJ_{n+1}(ne)-J_{n+2}(ne)\big]^2 \ + (1-e^2) \times \\
[J_{n-2}(ne) & - 2J_n(ne)+J_{n+2}(ne)]^2 +\frac{4}{3n^2}{[J_n(ne)]}^2\bigg) \, ,
\label{eq:harmonics}
\end{split}
\end{equation}
where $J_n$ denotes the Bessel function of first kind.
The above equations imply that the average radiated power is a rapidly rising function of the eccentricity (Peters \& Mathews 1963). It is also  worth noticing that whether a binary system has a non-zero eccentricity and  the emission of GWs is  strong,  its motion rapidly degenerate into a circular orbit. This happens  because the average value $ <de/dt>$ is negative in virtue of the radiation reaction. 

In what follows, we are interested only in exoplanetary systems with very short periods which potentially may produce reasonably intense GW emission. So,  let us  approximate  (\ref{LGW}) by taking $e=0$ to calculate the GW emission. In practise, this means that we have a lower bound on the absolute GW luminosity. In this case, the frequency of the emitted GW, $f_{GW} = \frac{2}{P}= \frac{1}{\pi}\sqrt{\frac{G(m_1+m_2)}{a^3}}$, is twice the orbital frequency.
Another important quantity when studying gravitational waves is the strain, the perturbation in the space-time metric. To obtain it, let us consider two point masses $m_1$ and $m_2$, but due to our approximation we consider again only circular orbits.  In the quadrupole approximation (Landau \& Lifshitz 1985),  the GW amplitudes are fully determined by the so-called  ``chirp mass'' of the binary system, $M_{\ast} \equiv\mu^{3/5}m^{2/5}$, where $\mu = m_1m_2/(m_1 + m_2)$ is the reduced mass and $m = m_1 + m_2$ is the total mass.   

After averaging over the orbital characteristic periods, the standard expression for the GW amplitude is readily obtained (Roelofs {\it{et al.}} 2006; Postnov \& Yungelson 2014):
\begin{eqnarray}
h = \left(\frac{32}{5}\right)^{1/2}\frac{G^{5/3}}{c^4}\frac{M_{\ast}^{5/3}}{d}(\pi f_{GW})^{2/3} \sqrt{cos^{4}i + 6cos^{2}i + 1} 
\label{eq02}
\end{eqnarray}
where $i$ is the inclination of the binary system, the angle between the line of sight to the binary orbit. Usually, the inclination effect is small but always increases the strain. 

Other key quantity when studying GW from binary systems is the period deviation or, more specifically, the cumulative period variation. By considering $e=0$, such an expression is easily obtained: 
\begin{eqnarray}
\frac{\dot P}{P}=-\frac{96}{5}\frac{G^{5/3}\mu m^{2/3}}{c^5}   \left( \frac{P}{2\pi}\right)^{-8/3}\,,   \label{eq03}
\end{eqnarray} 
where the cumulative shift of periastron time expression is
\begin{eqnarray}
\Delta Periastron = \frac{\dot P}{2P} (\Delta t)^2\,,\label{eq04}
\end{eqnarray}
where $\Delta t$ is the time of periastron passage (Maggiori 2008).

Given these well known results, let us now consider a specific sample of short period exoplanetary systems which can be seen as promising targets for a direct GW detection. 

\begin{table}
\centering
 \caption{Orbital quantities of 14 confirmed short period exoplanetary systems.  The data were taken from http://exoplanet.eu/catalog/. Due to their short period, the emitted gravitational power can be measured by the next generation of detectors (see next section).}\label{tab1}
\begin{tabular} {lccccc} \hline
\centering
Planet & $\frac{M_p}{M_J}$ & $\frac{M_s}{M_{\odot}}$ & $P$(day) & $a(10^5km)$ & $d(pc)$ \\  \hline
GP Com b &	26.2 & 0.435	& 0.032 &	2.28 & 75 \\
V396 Hya b & 18.3 & 0.345 &	0.045 &	2.64 &	77 \\
J1433 b	& 57.1 & 0.8 &	0.054 &	3.97 &	226 \\
PSR J1807-2459b &	9.4 & 1.4 & 	0.070 &	5.58 &	2790 \\
PSR 1719-14b &	1 & 1.4 &	0.090 &	6.58 &	1200 \\
PSR J2051-0827b &	28.3 & 1.4 &	0.099	& 7.05 &	1280 \\
PSR J2241-5236b &	12 & 1.35 &	0.145 &	8.96 &	500 \\
Kepler-70 b	& 0.014 & 0.496 &	0.240 &	8.95 &	1180 \\
Kepler-70 c	& 0.002 & 0.496 &	0.342 &	11.3 &	1180 \\
Kepler-78 b &	0.005 & 0.81 &	0.355 &	13.7 &	12.64 \\
PSR B1957+20b &	22 & 1.4 & 	0.380 &	17.3 &	1530 \\
EPIC 201637175b &	1.4 & 0.6 &	0.381 &	13.0 &	225 \\
CVSO 30 b &	6.2 & 0.39 &	0.448 &	12.6 &	330 \\
Kepler-42 c &	0.006 & 0.13 &	0.453 &	8.75 &	38.7 \\ \hline
\end{tabular}
\end{table}



\section{Exoplanetary (Short Period) Data} 

Surely, in comparison with black holes and neutron star binaries,  one may expect that exoplanetary systems are relatively weak sources of gravitational waves. However, since they are very near to us, in principle,   there may exist a special class of orbiting exoplanets that ripples the space-time beyond the expected stochastic background of GWs whose spectrum peaks at $10^{-5}$Hz (Ain, Kastha \& Mitra 2015) thereby providing favorable conditions for direct detection. 

\begin{table}
\centering
\caption{Calculated gravitational wave parameters of 3  exoplanetary systems  of extremely short periods  which are within the standard LISA sensitivity curve (see next section). The luminosity,  $L_{GW}$, is given in erg s$^-1$ and $\dot{P}/{P}$ in s$^{-1}$. In the last column we see the value of the inclination needed to obtain the strain $h$ according to equation (4). }\label{tab2}
\begin{tabular} {lcccc} \hline
Planet &  $L_{GW}(10^{30})$ & $h (10^{-22})$ & $\frac{\dot{P}}{P} (10^{-17})$ & $i (rad)$  \\ \hline
GP Com b  & $1.45$	& $1.98$&	$-3.44$ & $0.97$ \\
V396 Hya b & $0.17$ &	$0.98$ & $-0.83$ & $0.91$ \\ 
J1433 b & $2.69$&	$0.89$ & $-2.78$ & $1.47$ \\\hline
\end{tabular}
\end{table}

\begin{figure*}
\centerline{\epsfig{figure=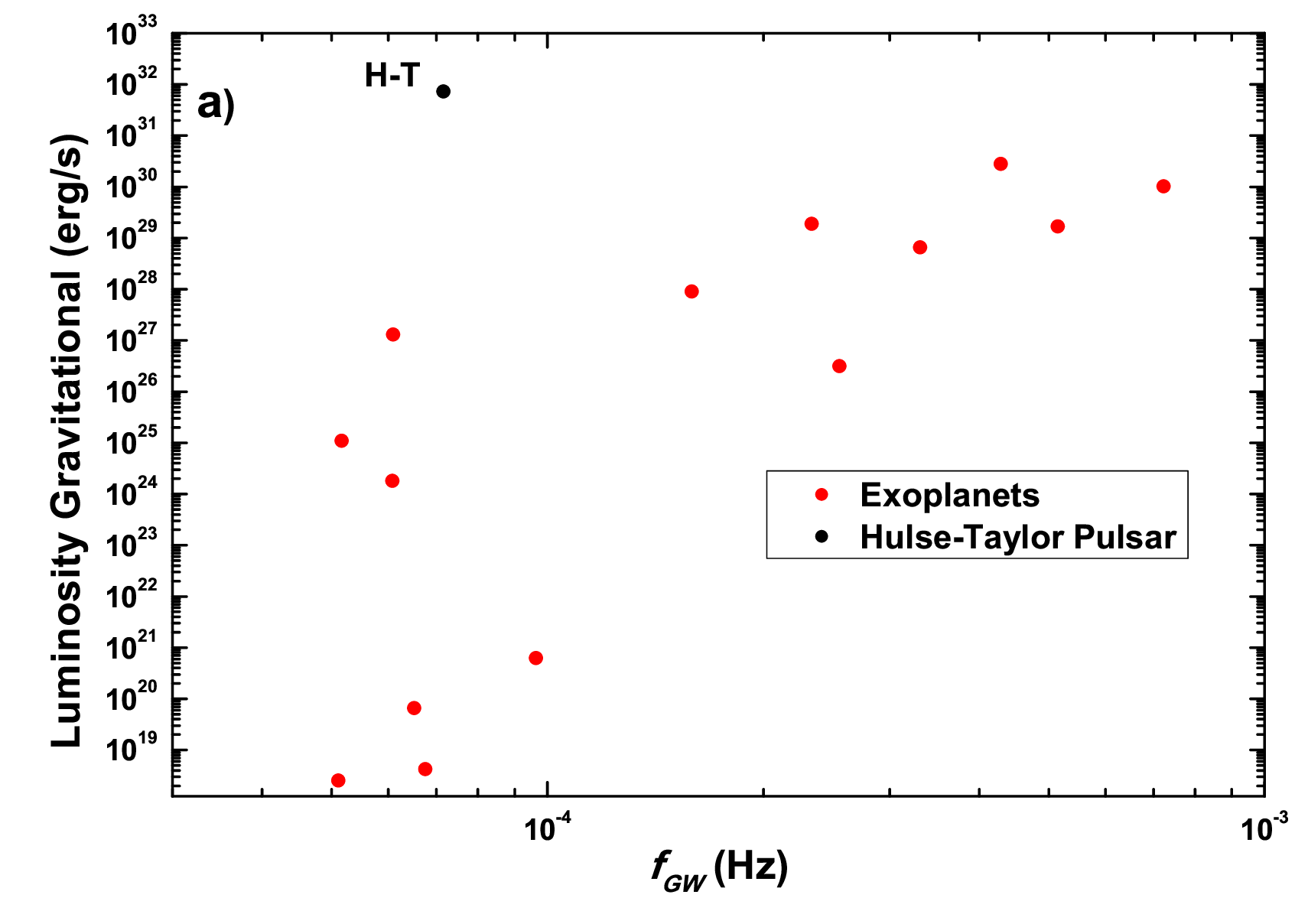,width=3.1truein,height=2.5truein}
++\epsfig{figure=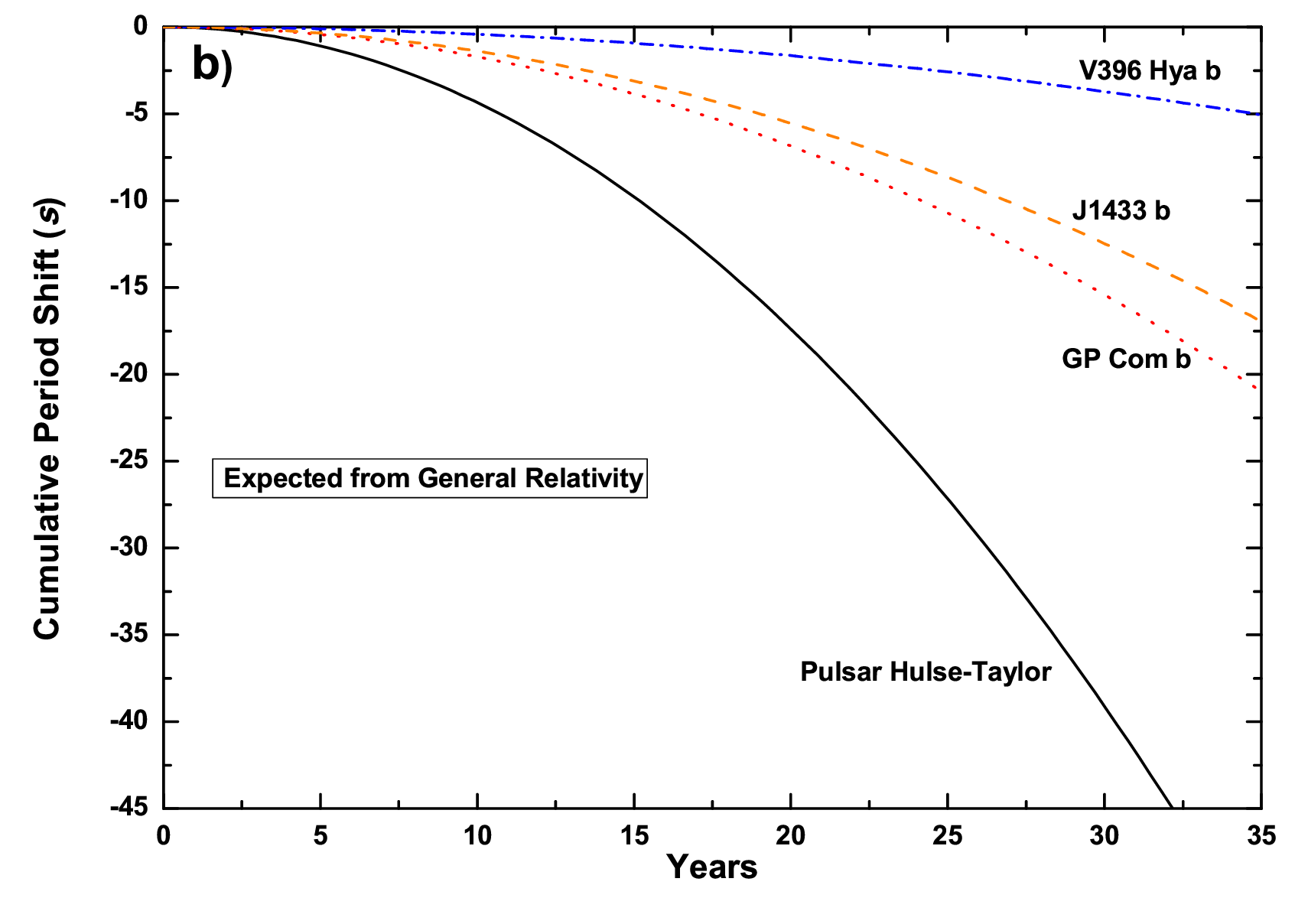,width=3.1truein,height=2.5truein}}
\begin{minipage}[t]{6in} \caption{a) Absolute GW luminosity as a function of the frequency $f_{GW}$. The black dot depict the PSR 1913+16 (H-T) pulsar while the 14 red dots are exoplanetary systems with known keplerian parameters (see Table 1). b) Precession of the periastron. By taking the errors on the orbital parameters we plot the theoretical prediction for the cumulative period shift of the 3 exoplanets listed in Table  2. For comparison we show  the standard  result for the H-T pulsar (solid black curve).} 
\end{minipage}
\end{figure*}

Keeping this in mind, we see from (1) and (5) that systems with  greatest masses and  smallest semi-major axis - those with shorter periods  ($P^2 = 4 \pi^2 a^ 3 / G m) $ - present the most favorable conditions to maximize the GWs emission.

In {Table  1}  we list the chosen exoplanets and their basic parameters. As one can see there, the most important orbital data for our calculations are: the mass of the planet in terms of the mass of Jupiter ($M_p/M_J$); the mass of the parent star in terms of the mass of the Sun ($ M_s/M_{\odot}$); the orbital period ($P$); the distance of the planet to the star ($a$); and the distance from the system to Earth ($d$). It has  been organized in descending order of the orbital period. It should also be noticed  that the periods of all exoplanets are restricted  on the interval $0.032 \leq P  (days) \leq 0.453$. Note also that some planets have masses equivalents to Earth ($M_E\simeq 0.00315 M_J$), and other solar planets.

In  {Table  2} we have selected a subset containing the most promising GW emitters formed by exoplanets and their parent stars based on the above criterion (shorter periods, see also {Table 1}).

The associated parameters in {Table 2} are: The calculated $L_{GW}$ and strain $h$ appearing in the first and second columns, respectively. In the third column we depicted the time variation of the orbital period, and, finally, in the fourth column, we show the orbital inclination angle of the selected subsystem ($i$). As we shall see, this last information is also interesting  because it represents for exoplanets a severe restriction to some  optical methods when compared with the detection by gravitational waves (see section 4). 

As far as we know, such objects are endowed with the most favorable  orbital  parameters whose GW pattern is detectable at least by two distinct instruments: (i) indirectly through the SKA measurements of the cumulative orbital phase shift of the periastron time, and (ii) directly by the space-based  LISA (see section 4). 

Observations and/or discovery of exoplanets is a very active field of research, and thus the number of objects is expected to climb substantially over the next few years. In this way, even more extreme systems, those with much shorter periods, may be discovered. 

It should also be noticed  that at least for the J1433 b system the electromagnetic luminosity, $(3.1 \pm 0.1)\times 10^{-4}L_{\odot} \simeq 1.18\times 10^{30}$ erg s$^{-1}$ (Hern\'andez Santisteban {\it{et al.}} 2016; Showman 2016), is nearly two times smaller than  the GW emission, $\simeq 2.69\times 10^{30}$ erg s$^{-1}$. Potentially, a GW emission at this level  may alter significantly the planetary habitable zone for this sort of systems. As we shall see next, these ultra short period exoplanetary probes, mainly binary systems like the ones shown in Table 2,  are  also suitable targets for the next generation of GWs detectors. Another interesting possibility  is the discovery of exoplanets of ultra short periods through gravitational waves (see next section).

\section{Discussion}

It is widely known that Hulse and Taylor indirectly measured GWs from the binary radio pulsar PSR 1913+16 and tested the predictions of general relativity based on its cumulative deviation of the periastron.  We argue here that the same can also be done for a subset of exoplanets, starting by the three short period systems listed in {Table 2}. 

In {Figure 1a} we display the calculated GW luminosity  as a function of the frequency for the 14 binary systems (red-points) appearing in Table 1.  For comparison we have also shown (see black dot) the Hulse-Taylor pulsar. 

To see how the distance is crucial, we recall that the binary Hulse-Taylor (H-T) pulsar emits $\sim 7.5\times 10^{31} erg s^{-1}$, it is at $8.4$ kpc and has cumulative period shift of the periastron of $45$ seconds in 33 years (Weisberg, Nice \& Taylor 2010). The three proposed targets have emission $\sim 10^{30} erg s^{-1}$ and distances of the order $75 - 226$ pc while their cumulative period deviation fall between $5 - 20$ seconds (see Table 2 and Figure 1b). In general,  the ratio between the apparent GW luminosities of exoplanet systems (Es) and other binary systems [e.g., pulsar systems (Ps)] arriving at the Earth surface reads:
\begin{equation}
\frac{\ell^{Es}}{\ell^{Ps}} = \frac{L^{Es}_{GW}}{L^{Ps}_{GW}}\left[\frac{d^{Ps}}{d^{Es}}\right]^{2}\,.
\end{equation}
where $d^{Es}$ and $d^{Ps}$ are the corresponding distances. This result follows directly from the standard $d^{-2}$ distance law (see section 2). In particular, for the  exoplanets of Table 2 and the H-T pulsar, the  corresponding ratios are, respectively: 242, 27 and 50. Thus, the GW flux on Earth surface from short period exoplanetary systems can be much greater than the one from H-T pulsar.

In  {Figure 1b} we show the theoretical prediction for  the cumulative deviation of the periastron by taking into account the orbital parameters (objects for lines blue, green and red are indicated in the insert). The black line stands for the  well known H-T pulsar. By comparing with the three predicted curves for the exoplanets with the binary radio H-T pulsar  one may see the interest to observe this subset of exoplanets with the same techniques. In principle, such objects can be investigated through a combination of radio and optical observations, as usually done for white dwarfs binaries and pulsars (Taylor \& Weisberg 1989; Antoniadis {\it{et al.}} 2013). 

In  {Figure  2} we display the calculated  strain as predicted  from  orbital parameters of the exoplanets versus the GW frequency. The black line represents the LISA sensitivity curve (Larson \& Hiscock 2000; Prince, Tinto \& Larson 2002), while the red circles are the exoplanetary systems of Table 1. The LISA sensitivity curve was obtained using the typical data: signal to noise ratio $1.0$, arm length $5.0\times 10^{9}$ meters, laser power $1.0$W, etc. - more details in Larson's personal page http://www.srl.caltech.edu/$\sim$shane/sensitivity/ - in this connection see also also Petiteau et al. (2008). For comparison we also show the strain associated to the H-T Pulsar. Clearly, one may conclude that the objects of Table 2 can be detected by the space-based LISA instrument. 

It thus follows that short period exoplanetary systems can be considered as a distinct subset of binaries sources forming suitable targets for prospecting gravitational waves. In principle, the possibility of detection must be enhanced for the next generation of space-based and ground-based instruments. 

Now, let us finish this section  drawing attention to some interesting  physical effects  of  gravitational waves  from exoplanetary systems of ultra short period to different areas: 

\begin{itemize}

\item  The planetary habitable zone is, primarily, the distance to the parent star where liquid water may exist. Thus, exoplanetary systems whose GW luminosity  is of the same order or greater than the electromagnetic luminosity from the central star  (see the specific example in Section 3) may change significantly the planetary habitable zones as long as there is a mechanism converting gravitational radiation into heat.  Corrections of the order of $(L_{GW}/L_{EM})^{1/2}$ where $L_{EM}$ is the electromagnetic luminosity, are expected. This could be the first direct connection involving general relativistic effects, habitable zone and the possibility of formation and maintenance of life. The physical consequences of GW for circumstellar habitable zones deserves a closer scrutiny and it will be detailed investigated in a forthcoming communication.

\item To date many multiple exoplanetary systems were already discovered (see, for instance, Luger {\it et al.} 2017). The presence of any inner short period planet will change not only the local habitable zone but also creates new interesting possibilities.  The remaining planets behave like  gravitational antennas by absorbing and transforming in heat part of the incident GWs energy. This is interesting because the energy changes in  planets are well modeled, especially Earth-like planets whose electromagnetic environmental interaction has been discussed for decades (Coughlin \& Harms 2014; Lopes \& Silk 2015; Mulargia 2017). Any observed anomalous thermal behavior  may impose constraints on the cross section and also in the interaction between matter and the incident gravitational waves. In principle,  this kind of observation could be used to test new aspects of general relativity and other gravity theories. 

\item Detection of exoplanets by optical techniques is a time-consuming task. For transiting time, for instance, this happens because not all planet's orbit as seen from the Earth  have the precise geometry to eclipse, and, as such, the probability to obtain the right narrow range of inclinations is very low (the transit probability to the Earth  viewed outside from Solar System is less than 1\%). Actually, precise transit time measurements require nearly edge-on planetary orbits observable only by a  special class of observers. Apart measurements with enough precision, time sampling and spectroscopic follow-up observations, the low probability also  explains why a great number of stars must be searched before expecting to find a transiting planet. However, this is not the case for a direct detection through gravitational waves. For ultra short period systems like the ones appearing in Table 2, the detection of gravitational waves may provide an interesting tool for discovering new exoplanets, in principle,  even more efficient than the available optical methods.    


All these connections reinforce the idea that nearby short period  exoplanetary systems provide a key  class of GW emitters which may be relevant both from a methodological and a physical viewpoint. Potentially, as argued here, such systems are capable to open new avenues of investigation.   

\end{itemize}
\begin{figure}
\centerline{\epsfig{figure=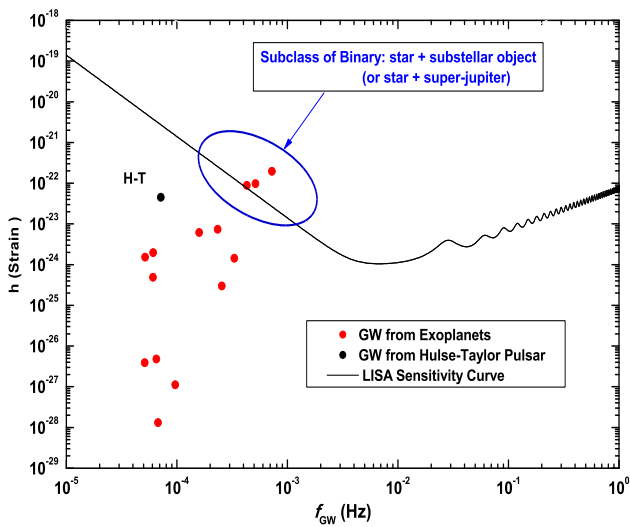,width=3.4truein,height=2.8truein}}
\caption{Strain versus frequency for the exoplanets listed in Table 1. Black solid line is a characteristic LISA sensitivity curve for one year of integration and typical parameters as mentioned in the text (see Larson {\it et al.} 2000). Note that the 3 objects within the blue elipse also appearing in Table 2 may be detected by LISA.}
\end{figure}

\section{Conclusions}

Based on some already observed sources we have shown that a subset of ultra short period exoplanetary systems are suitable targets to direct detection of gravitational waves for the next generation (space-based) instruments (see {Table 2} and {Figure 2}). In principle, as shown in ({Figure 1b}), even indirect observations  of gravitational waves through  the cumulative  time variation of the periastron for this sort of extrasolar planets should  be considered an interesting possibility.  

The examples discussed here suggest the existence of a subset of nearby exoplanetary systems  composed by relatively massive bodies endowed with small orbital period thereby making the GW emission (and the associated strain) strong enough to be detected from Earth. Although less intense than extreme binary systems, the gravitational wave emission of ultra short period exoplanets are interesting targets for the next generation of planned instruments. 

Some interesting observational consequences are: (i) the  gravitational wave emission must change the electromagnetic habitable zone around multi-exoplanetary systems; (ii) limits on the absorption cross section of the gravitational wave interacting with matter can be constrained by the outer planets working like gravitational  antennas  absorbing and decreasing the original signal, and (iii) the emitted  gravitational wave pattern may provide an efficient tool to discover new ultra short period exoplanets. Hopefully,  studies in this field may bring valuable results to different areas including astrobiology, physics and astrophysics. 

\section*{Acknowledgements}
The authors thank L. A. Almeida, J. D. Nascimento and  C. E. Pellicer, for helpful discussions. JASL is grateful to CAPES (PROCAD project, 2013), FAPESP (LLAMA Project) and fellowship from CNPq. The basic ideas of this paper were discussed during the 3rd Workshop on Cosmology and Gravitation: Celebrating the 100 years of Cosmology (1917- 2017) in Natal-RN, Brazil.

\bsp
\label{lastpage}
\end{document}